\documentclass[aps,prd,twocolumn,superscriptaddress,amsmath,showpacs]{revtex4-1}
\usepackage{graphicx}
\usepackage{dcolumn}
\usepackage{bm}
\usepackage{natbib}
\newcommand{\be}{\begin{equation}}
\newcommand{\ee}{\end{equation}}
\newcommand{\bea}{\begin{eqnarray}}
\newcommand{\eea}{\end{eqnarray}}

\newcommand{\Mpc}{{\rm ~Mpc}}

\urldef\mgcamb\url{http://www.sfu.ca/~aha25/MGCAMB.html}

\begin{document}
%opening
\title{Dark Radiation after Planck}
\author{Najla Said}
\affiliation{Physics Department and INFN, Universit\`a di Roma ``La Sapienza'', Ple Aldo Moro 2, 00185, Rome, Italy}

\author{Eleonora Di Valentino}
\affiliation{Physics Department and INFN, Universit\`a di Roma ``La Sapienza'', Ple Aldo Moro 2, 00185, Rome, Italy}

\author{Martina Gerbino}
\affiliation{Physics Department and INFN, Universit\`a di Roma ``La Sapienza'', Ple Aldo Moro 2, 00185, Rome, Italy}

\begin{abstract}
We present new constraints on the relativistic neutrino effective number $N_{\rm eff}$ and on the
Cosmic Microwave Background power spectrum lensing amplitude $A_{\rm L}$ from the recent Planck 2013 data release. 
Including observations of the CMB large angular scale polarization from the WMAP satellite, 
we obtain the bounds $N_{\rm eff}=3.71\pm0.40$ and $A_{\rm L}=1.25\pm0.13$ at $68 \%$ c.l..
The Planck dataset alone is therefore suggesting the presence of a dark radiation component
at $91.1\%$ c.l. and hinting for a higher power spectrum lensing amplitude
at $94.3\%$ c.l.. 
We discuss the agreement of these results with the previous constraints obtained from the Atacama Cosmology Telescope (ACT) and the South Pole Telescope (SPT). Considering the constraints on the cosmological parameters, we found a very good agreement with the previous 
WMAP+SPT analysis but a tension with the WMAP+ACT results, with the only exception of the lensing amplitude. 
\end{abstract}


\maketitle

\section{Introduction} \label {sec:intro}

The recent precise measurements of the Cosmic Microwave Background (CMB hereafter) temperature anisotropies 
released by the Planck collaboration \cite{Ade:2013xsa} 
are providing the tightest constraints on cosmological parameters to date \cite{Ade:2013lta}.

In this paper, we use this new dataset to constrain two parameters that affect the
CMB ``damping tail" regime, at small angular scales, namely the
neutrino effective number $N_{\rm eff}$ and the lensing amplitude $A_{\rm L}$, that,
from previous experiments, have been reported as not consistent with the standard expectations \cite{DiValentino:2013mt}.

We remind here that $N_{\rm eff}$ effectively measures the number of relativistic 
degrees of freedom at recombination and is related to the energy density
in relativistic ``dark" particles $\rho_{\nu}$ by:

\begin{equation}
\rho_\nu=   \left[\frac{7}{8}\, \left( \frac{4}{11} \right)^\frac{4}{3} N_{\rm eff}\right] \rho_\gamma\, , 
\label{eq:neffdef} 
\end{equation} 
where $\rho_\gamma$ is the CMB photon energy density, with value today
$\rho_{\gamma,0}\approx 4.8 \times 10^{-34}$ g\,cm$^{-3}$.

In the standard scenario, assuming three relativistic neutrino families,
the expected value is $N_{\rm eff}=3.046$. Observation of a different value could
point to new physics, related to the neutrino sector, such as non standard
neutrino decoupling, sterile neutrinos, etc., or to even more exotic physics, such as axions, extra dimensions, early dark energy (see e.g. \cite{newphys}, \cite{DiValentino:2013qma}
and references therein).

On the other hand, the $A_{\rm L}$ parameter is a phenomenological parameter introduced in \cite{Calabrese:2008rt}, 
that simply rescales the lensing potential:
\begin{equation}
C_{\ell}^{\phi \phi} \rightarrow A_{\rm L} C_{\ell}^{\phi \phi}
\end{equation}

\noindent where $C_{\ell}^{\phi \phi}$ is the power spectrum of the
lensing field.
The expected value for this parameter in the standard framework is
$A_{\rm L}=1$. A value different from one could indicate either the presence of
a systematic, or the presence of new physics (see e.g.
\cite{DiValentino:2013mt} and \cite{marchini}).

The previous CMB measurements obtained by the Atacama Cosmology Telescope (ACT, \cite{act2013}) and the
South Pole Telescope (SPT, \cite{spt2013}), when combined with the
latest observations from the WMAP satellite (WMAP9, \cite{Bennett:2012fp}), have indeed provided different values 
for these two parameters, that are in tension at the level of two standard deviations.
As showed in \cite{DiValentino:2013mt}, the ACT dataset gives $N_{\rm eff}=2.85\pm0.56$ and $A_{\rm L}=1.64\pm0.36$ 
at $68 \%$ c.l., while the SPT dataset gives $N_{\rm eff}=3.72\pm0.46$ and $A_{\rm L}=0.85\pm0.13$ at $68 \%$ c.l..

Given this tension, it is certainly timely to investigate the constraints that
can be obtained for $N_{\rm eff}$ and $A_{\rm L}$ from the new data from the Planck
satellite. 

The Planck collaboration has already presented results in
\cite{Ade:2013lta} on $N_{\rm eff}$ and $A_{\rm L}$ separately.
Here we extend this analysis by varying $N_{\rm eff}$ and $A_{\rm L}$ simultaneously,
i.e. taking into account the possible correlations between these
two parameters as in \cite{DiValentino:2013mt}, and by properly comparing the results with the
previous ACT and SPT measurements in the $N_{\rm eff}$-$A_{\rm L}$ plane. 

Our paper is simply organized as follows: in the next section we describe
the analysis method, in Section III we present our results also considering Baryon Acoustic
Oscillation (BAO) surveys and $H_0$ measurements, while 
in Section IV we derive our conclusions.

\section{Data analysis method}\label{sec:data}

\begin{table*}
\begin{center}
\begin{tabular}{|c|c|c|c|}
\hline\hline
Parameter & Planck+WP & WMAP9+SPT & WMAP9+ACT \\
\hline
$\Omega_bh^2$ &$0.02306\pm0.00051$ &$0.02264\pm0.00051$ &$0.02283\pm0.00052$ \\
$\Omega_{\rm c}h^2$ &$0.1239\pm0.0054$ &$0.1232\pm0.0080$ &$0.110\pm0.010$\\
$\theta$ &$1.04124\pm0.00077$ &$1.0415\pm0.0012$ &$1.0412\pm0.0025$ \\
$\tau$ &$0.095\pm0.015$ &$0.088\pm0.014$ &$0.090\pm0.014$\\
$n_s$ &$0.996\pm0.018$ &$0.982\pm0.018$ &$0.969\pm0.019$ \\
$log[10^{10} A_s]$ &$3.111\pm0.034$ &$3.169\pm0.048$ &$3.174\pm0.045$\\
$N_{\rm eff}$ &$3.71\pm0.40$ &$3.72\pm0.46$ &$2.85\pm0.56$\\
$A_{\rm L}$ &$1.25\pm0.13$ &$0.85\pm0.13$ &$1.64\pm0.36$\\
\hline
$\Omega_{\Lambda}$ &$0.736\pm0.022$ &$0.736\pm0.023$ &$0.728\pm0.025$\\
$t_0 [\mathrm{Gyr}]$ &$13.08\pm0.38$ &$13.14\pm0.43$ &$13.90\pm0.55$\\
$\Omega_m$ &$0.264\pm0.022$ &$0.264\pm0.023$ &$0.272\pm0.025$ \\
$H_0 [\mathrm{km}/\mathrm{s}/\mathrm{Mpc}]$& $74.9\pm3.7$ &$74.6\pm3.7$ & $69.9\pm3.7$ \\
\hline\hline
 \end{tabular}
 \caption{Constraints at $68 \%$ confidence level on cosmological parameters from our analysis using Planck+WP, WMAP9+SPT and WMAP9+ACT.}
 \label{plancktable_cmb}
 \end{center}
 \end{table*}

\begin{figure*}[htb!]
\begin{center}
\includegraphics[width=8cm,height=8cm]{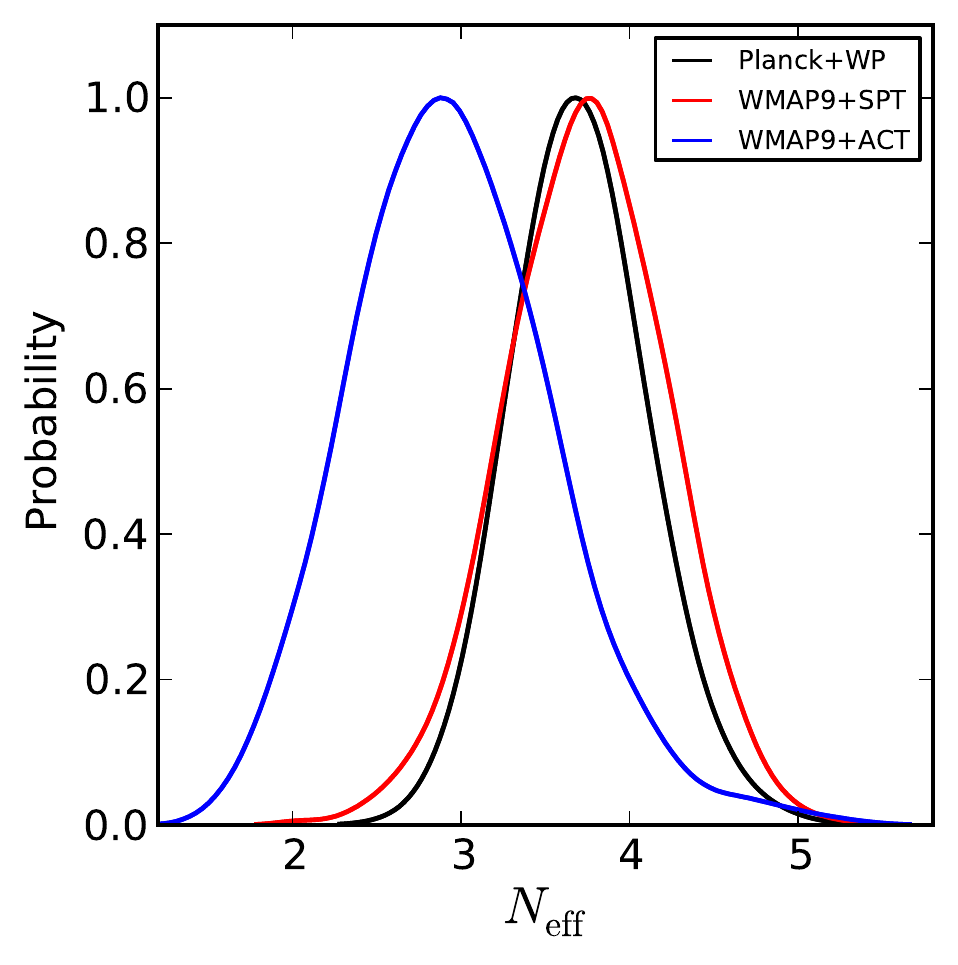} 
\includegraphics[width=8cm,height=8cm]{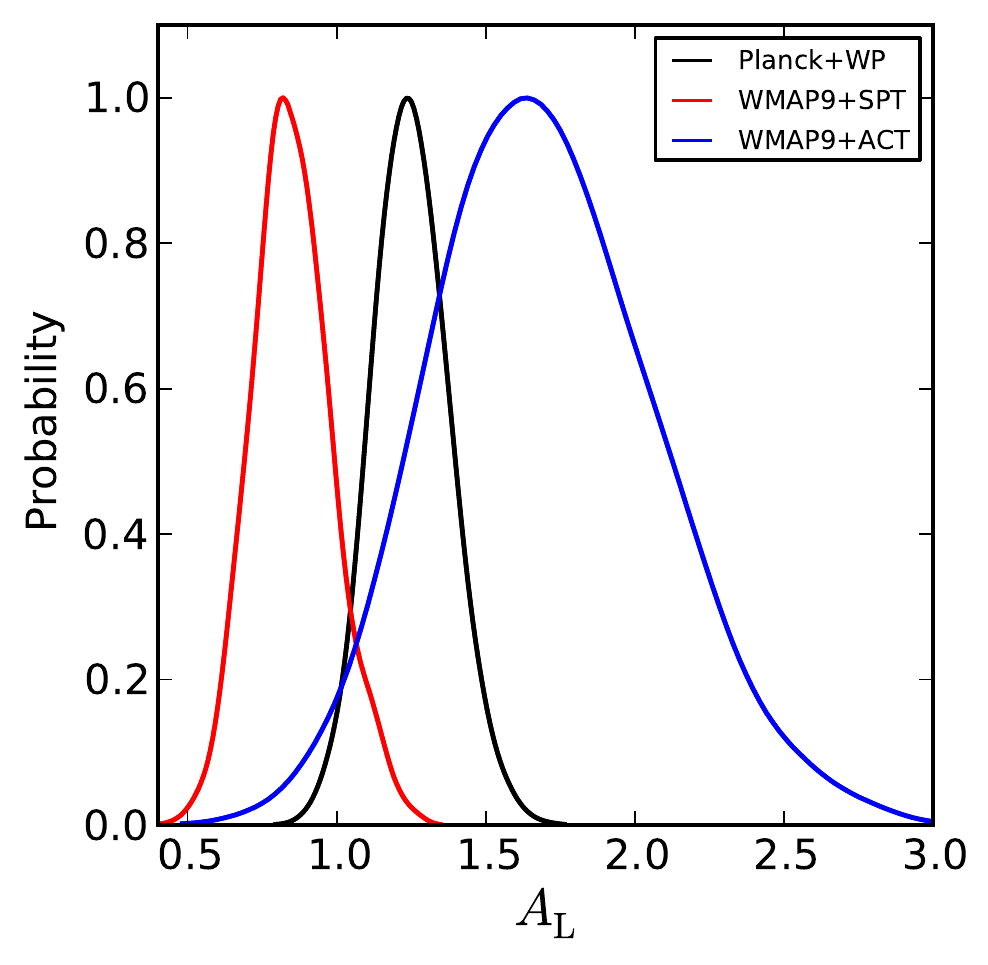} 
\caption{Comparison of the results for Planck+WP, WMAP9+SPT and WMAP9+ACT datasets in terms of the 1-D posterior distribution functions
for the parameters $N_{\rm eff}$ (left) and $A_{\rm L}$ (right).}
\label{1D_cmb}
\end{center}
\end{figure*}

\begin{figure}[htb!]
\begin{center}
\includegraphics[width=8cm]{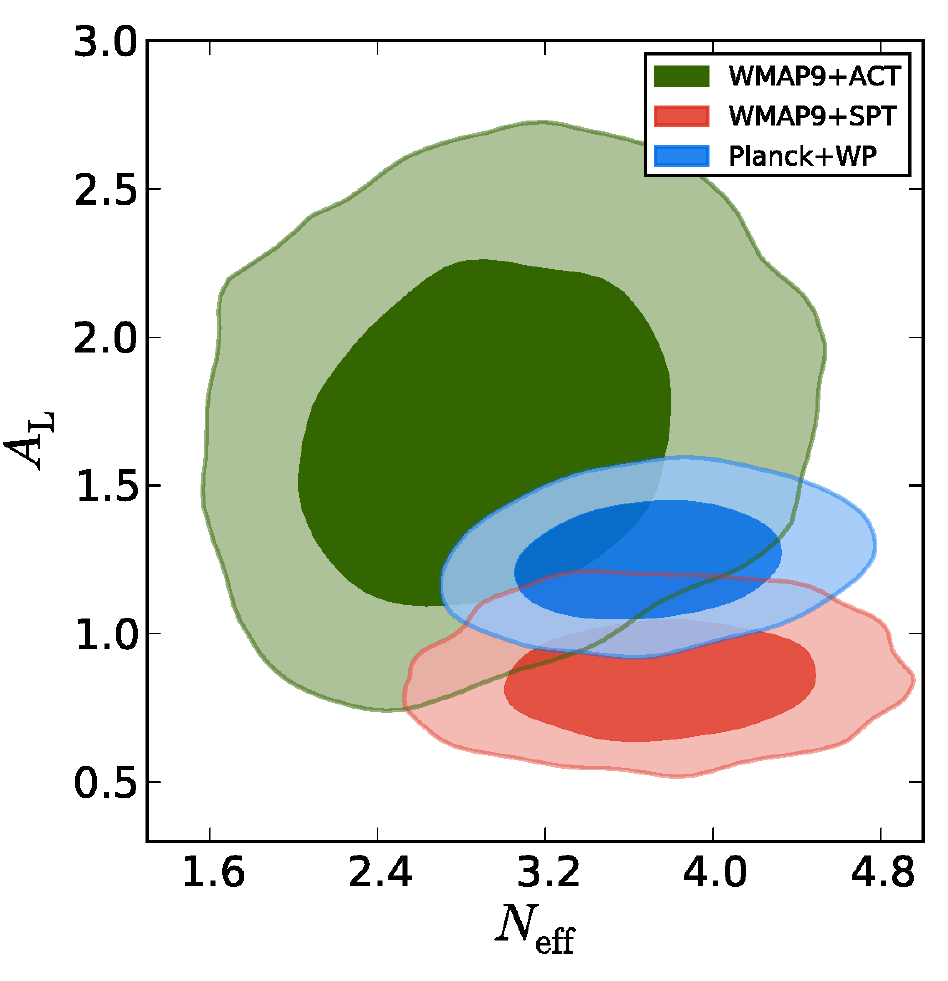}
\caption{Comparison of the 2-D posterior distribution function from the Planck+WP, WMAP9+ACT and WMAP9+SPT datasets in 
the $N_{\rm eff}-A_{\rm L}$ plane. }  
\label{nnu-alens_cmb}
\end{center}
\end{figure}

Our main CMB dataset consists in the Planck public data release of March 2013 \cite{Ade:2013xsa}. 
We compare this dataset with the theoretical models using the \texttt{CAMspec} likelihood version 6.2 for high multipoles and
the \texttt{commander} version 4.1 likelihood for low multipoles \cite{Planck:2013kta}. We also consider the WMAP low-$\ell$ likelihood for polarization \cite{Bennett:2012fp}. This dataset is identical to the ``PLANCK+WP" case presented in the Planck papers \cite{Planck:2013kta, Ade:2013lta}.

For BAO surveys we include the following datasets: SDSS-DR7 \cite{Padmanabhan:2012hf} at redshift $z=0.35$, 
SDSS-DR9 \cite{Anderson:2012sa} at $z=0.57$ and WiggleZ \cite{Blake:2011wn} at $z=0.44$, $0.60$, and $0.73$. 

Finally, we include the recent measurements for the Hubble constant $H_0$ from the analysis of \cite{Riess:2011yx} and
we refer to this dataset as HST.

For the analysis method we use the publicly available Monte Carlo Markov Chain package \texttt{cosmomc} \cite{Lewis:2002ah} which relies on a convergence diagnostic based on the Gelman and Rubin statistic. We use the latest version (March 2013) which includes the support for the Planck Likelihood Code v1.0 (see \url{http://cosmologist.info/cosmomc/}). The plots shown in this work are obtained via the \texttt{python} codes included in the \texttt{cosmomc} package.

We run over the six-dimensional space of standard cosmological parameters: the baryon and cold dark matter densities
$\Omega_{\rm b}$ and $\Omega_{\rm c}$, the ratio of the sound horizon 
to the angular diameter distance at decoupling $\theta$, the reionization optical depth $\tau$, the scalar spectral index $n_S$, and the overall normalization of the spectrum $A_S$ at $k=0.05\Mpc^{-1}$. We consider purely adiabatic initial conditions and
we impose spatial flatness. In addiction to these parameters we let the number of neutrinos species (assumed massless) $N_{\rm eff}$ and the lensing amplitude parameter $A_{\rm L}$ to vary, assuming the following flat priors: $1.047 \le N_{\rm eff} \le 10$ and  $0.0 \le A_{\rm L} \le 4.0$. 

In our runs, we also marginalise over the foreground parameters as in \cite{Planck:2013kta, Ade:2013lta}. Since
the correlations between the cosmological and foreground parameters is minimal, 
we do not report their values in this paper. The posteriors on foregrounds are in excellent agreement with those
reported in \cite{Ade:2013lta}.

\section{Results}\label{sec:results}

\subsection{CMB data only}
 
We first discuss our results obtained using the Planck+WP dataset. 
Posteriors on the cosmological parameters are shown in the first column of Table \ref{plancktable_cmb}. 

As we can see, the Planck+WP dataset provides an indication for a larger value
of  {\it both} $N_{\rm eff}$ and $A_{\rm L}$, with $N_{\rm eff}=3.71\pm0.40$ and $A_{\rm L}=1.25\pm0.13$ at
$68\%$ c.l..

The constraints on $N_{\rm eff}$ reported by the Planck collaboration for the same
dataset is $N_{\rm eff}=3.51\pm0.39$ at $68 \%$ c.l.. However, this constraint is obtained
with the condition $A_{\rm L}=1$. The slightly larger value obtained in our analysis clearly shows
that there is a small correlation between these two parameters and that fixing $A_{\rm L}=1$ could
slightly bias the constraints on $N_{\rm eff}$ to smaller values.

It is interesting to compare these results with the ones previously obtained in
\cite{DiValentino:2013mt} for the ACT and SPT datasets, as shown in Table \ref{plancktable_cmb}.
As we can see, the Planck+WP result on $N_{\rm eff}$ is perfectly consistent with the 
WMAP9+SPT constraint of $N_{\rm eff}=3.72\pm0.46$, while there is a tension with the WMAP9+SPT
result on $A_{\rm L}=0.85\pm0.13$.
Vice-versa, the WMAP9+ACT constraint $N_{\rm eff}=2.85\pm0.56$ is clearly in tension
with the Planck+WP result, while there is a better agreement with the bound
on the lensing parameter $A_{\rm L}=1.64\pm0.36$.

In order to better demonstrate the consistency between these datasets, we plot in Figure \ref{1D_cmb} 
the 1-D posterior distribution function for $A_{\rm L}$ and $N_{\rm eff}$ coming from these three
analyses, while in Figure \ref{nnu-alens_cmb} we report the constraints in the 2-D $N_{\rm eff}-A_{\rm L}$ plane. As we can see, the Planck+WP constraint on $N_{\rm eff}$ is in impressive agreement
with the previous WMAP9+SPT constraint and in tension with the WMAP9+ACT value.
On the other hand, the constraint on $A_{\rm L}$ from Planck+WP is in better agreement with the WMAP9+ACT result
but also consistent with the WMAP9+SPT constraint.

\begin{table*}
\begin{center}
\begin{tabular}{|c|c|c|c|}
\hline\hline
Parameter & CMB+HST & CMB+BAO & CMB+BAO+HST \\
\hline
$\Omega_bh^2$ &$0.022953\pm0.00035$ &$0.02246\pm0.00031$ &$0.02262\pm0.00028$ \\
$\Omega_{\rm c}h^2$ &$0.1234\pm0.0050$ &$0.1232\pm0.0053$ &$0.1260\pm0.0049$\\
$\theta$ &$1.04123\pm0.00077$ &$1.04112\pm0.00078$ &$1.04085\pm0.00075$ \\
$\tau$ &$0.094\pm0.014$ &$0.087\pm0.013$ &$0.089\pm0.013$\\
$n_s$ &$0.992\pm0.011$ &$0.974\pm0.011$ &$0.9815\pm0.0088$ \\
$log[10^{10} A_s]$ &$3.108\pm0.030$ &$3.093\pm0.030$ &$3.103\pm0.029$\\
$N_{\rm eff}$ &$3.63\pm0.27$ &$3.35\pm0.31$ &$3.56\pm0.27$\\
$A_{\rm L}$ &$1.24\pm0.12$ &$1.16\pm0.10$ &$1.17\pm0.10$\\
\hline
$\Omega_{\Lambda}$ &$0.733\pm0.014$ &$0.706\pm0.011$ &$0.7119\pm0.0094$\\
$t_0 [\mathrm{Gyr}]$ &$13.15\pm0.23$ &$13.47\pm0.28$ &$13.27\pm0.23$\\
$\Omega_m$ &$0.267\pm0.014$ &$0.294\pm0.011$ &$0.2881\pm0.0094$ \\
$H_0 [\mathrm{km}/\mathrm{s}/\mathrm{Mpc}]$& $74.0\pm2.0$ &$70.4\pm1.9$ & $71.8\pm1.6$ \\
\hline\hline
 \end{tabular}
 \caption{Constraints at $68 \%$ confidence level on cosmological parameters from our analysis using CMB+HST, CMB+BAO and CMB+BAO+HST.}
 \label{plancktable}
 \end{center}
 \end{table*}

\begin{figure*}[htb!]
\begin{center}
\includegraphics[width=8cm,height=8cm]{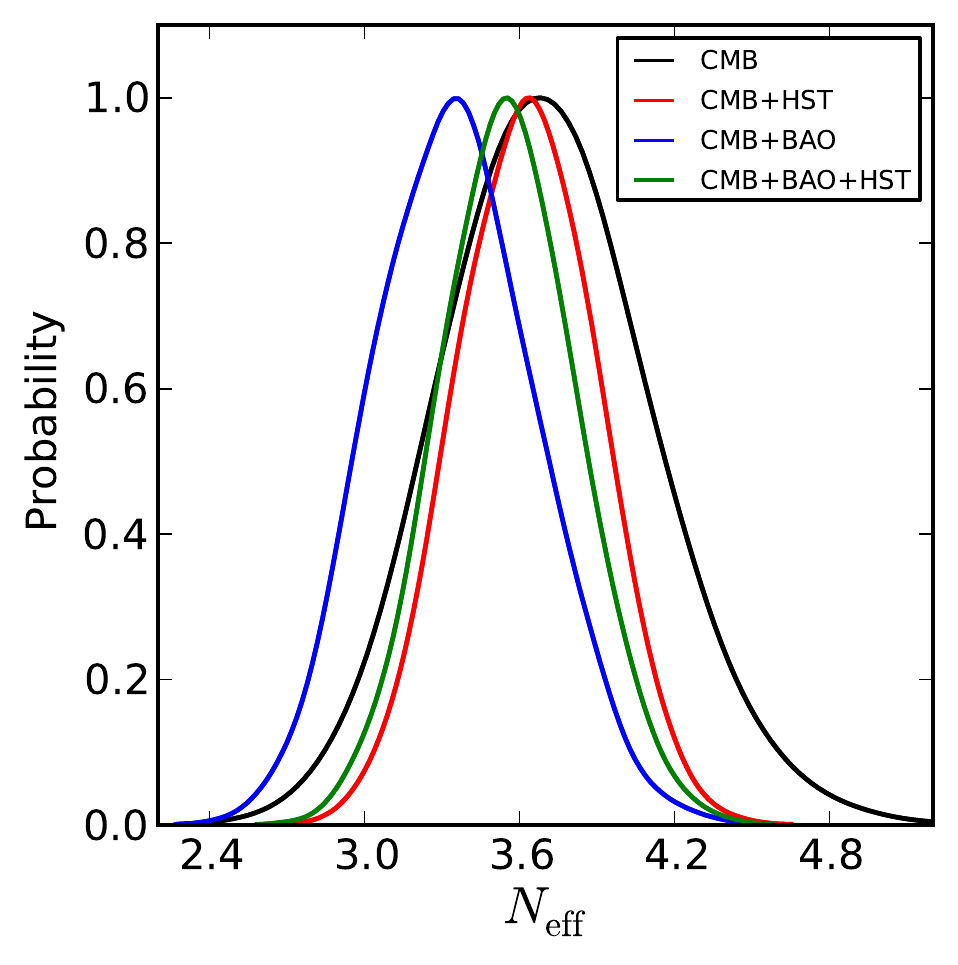} 
\includegraphics[width=8cm,height=8cm]{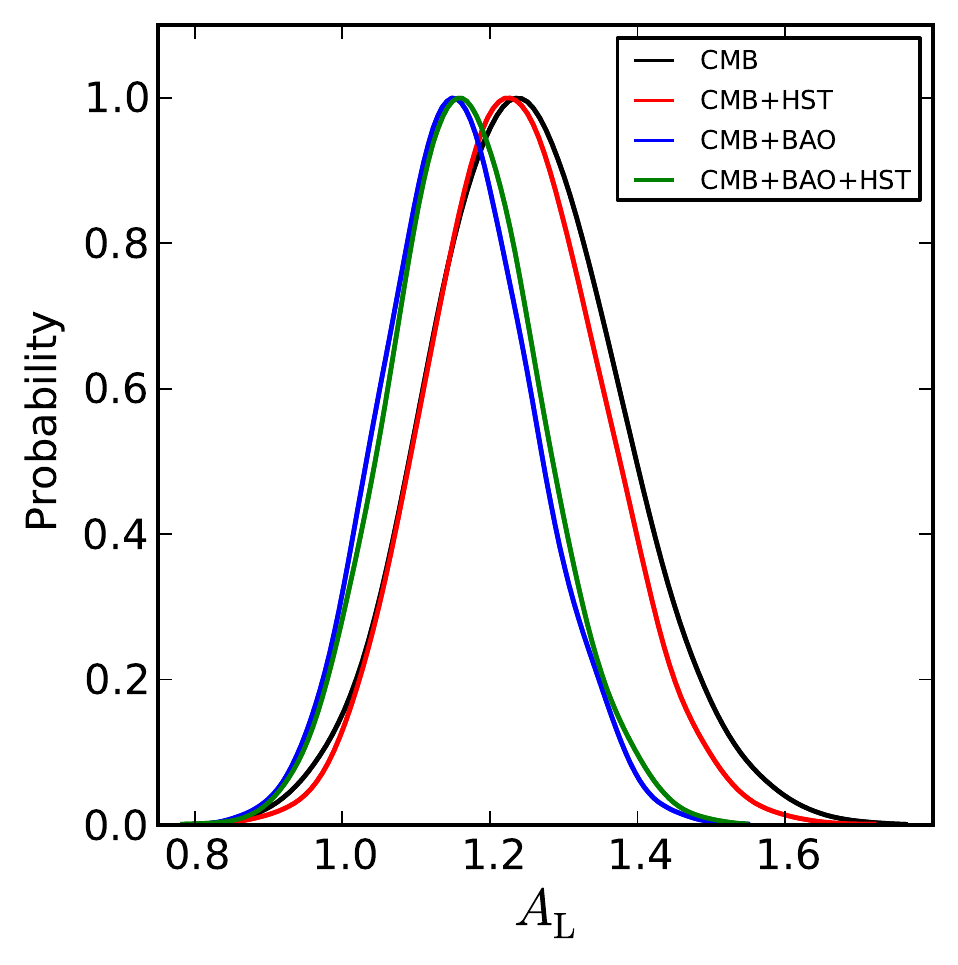} 
\caption{Comparison of the 1-D posterior distribution functions from the CMB-only (Planck+WP), CMB+HST, CMB+BAO and CMB+BAO+HST datasets for $N_{\rm eff}$ (left) and
$A_{\rm L}$ (right).}
\label{1D_planck}
\end{center}
\end{figure*}

\begin{figure}[htb!]
\includegraphics[width=8cm]{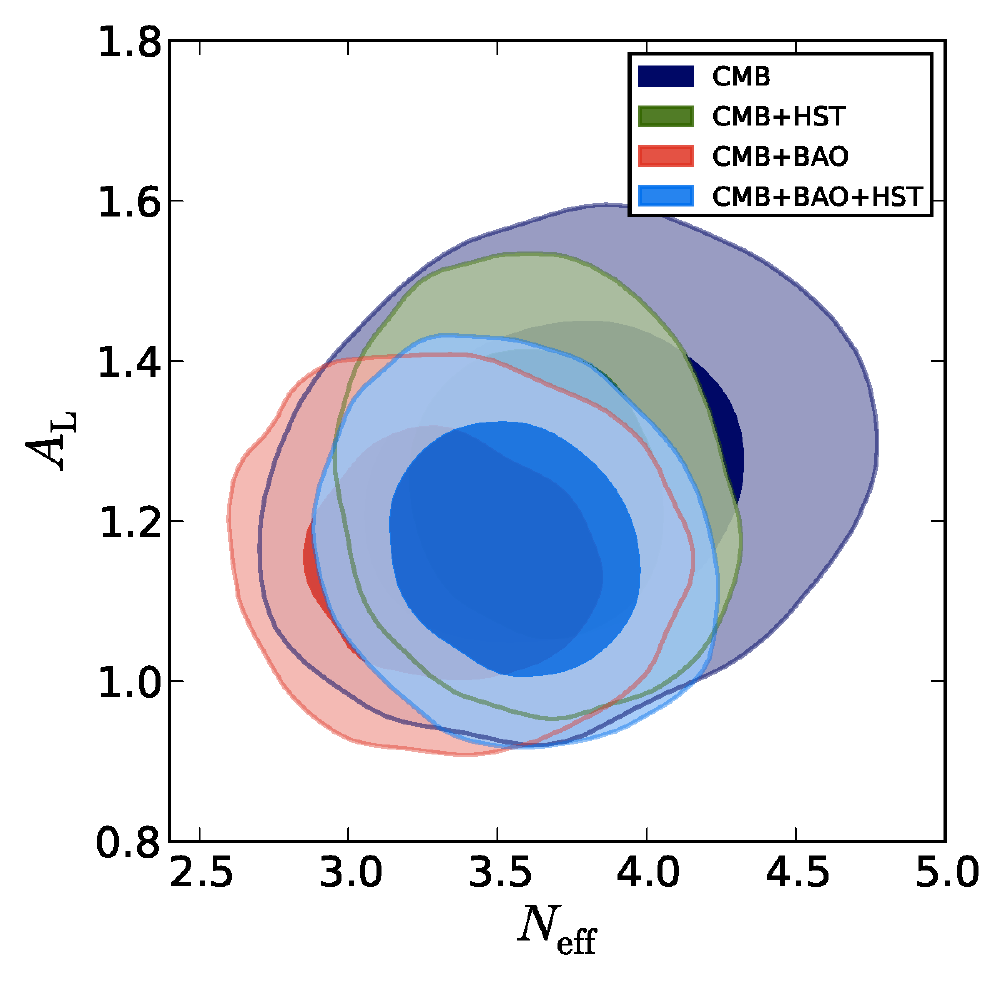}
\caption{Comparison of the 2-D posterior distribution functions from the CMB-only, CMB+HST, CMB+BAO and CMB+BAO+HST datasets
 in the $N_{\rm eff}-A_{\rm L}$ parameters plane. }  
\label{nnu-alens_planck}
\end{figure}

It is also interesting to note that the Planck+WP dataset is in general in better agreement
with the WMAP9+SPT dataset than the WMAP9+ACT on most of the parameters as the Hubble constant $H_0$, the matter density $\Omega_m$ and
the scalar spectral index $n_S$. This could appear as in contradiction with the results presented in
\cite{Ade:2013lta} where, on the contrary, a better agreement with the WMAP9+ACT dataset was
reported. The reason of this different conclusion is due to the fact that while in \cite{Ade:2013lta}
the comparison of Planck with ACT and SPT has been done by fixing $A_{\rm L}=1$, here we let this parameter
to vary freely. A similar conclusion has been reached by the SPT collaboration \cite{carlstrom}.

\subsection{Adding BAO and HST}

In this section we include the HST and BAO datasets.
We report the results obtained from the Planck+WP+HST, Planck+WP+BAO and
Planck+WP+HST+BAO analyses in the three columns of Table \ref{plancktable}, respectively.
For the sake of brevity, we refer to the Planck+WP dataset as ``CMB". 

We show the 1-D posterior probability distributions for $N_{\rm eff}$ and $A_{\rm L}$ in Figure \ref{1D_planck} and the 2-D confidence regions for $N_{\rm eff}$ and $A_{\rm L}$ in Figure \ref{nnu-alens_planck} for all the dataset combinations discussed: CMB+HST, CMB+BAO and CMB+BAO+HST. 

Our results are in perfect agreement with those already presented in \cite{Ade:2013lta}. In fact,
the introduction of the BAO dataset tends to drag down both the value of $N_{\rm eff}$ and $A_{\rm L}$, 
i.e. to a better consistency with the standard expectation; on the other hand, the inclusion of the HST 
dataset preserves the Planck+WP mean values of parameters, reducing the error bars and
therefore increasing the hints for new physics. In particular, for the CMB+HST case,
both $N_{\rm eff}$ and $A_{\rm L}$ assume larger values at more than $95 \%$ confidence level. 

When both BAO and HST are combined, the final effect is to lower the value of $A_{\rm L} = 1.17 \pm 0.10$, which however is still 
not in full agreement with the expected value of unity, and to lower the number of neutrinos species as well, giving $N_{\rm eff} = 3.56 \pm 0.27$, which also remains at almost $2\sigma$ away from the standard value.

\section{Conclusions}\label{sec:conclusions}

In this brief paper we have reported new joint constraints on the neutrino effective number $N_{\rm eff}$
and the CMB lensing amplitude $A_{\rm L}$ from the new Planck dataset.
We have shown that the Planck+WP dataset is hinting for both a presence of dark radiation
(at the level of $91.1\%$) and for an anomalous amplitude for the lensing
parameter (at the level of $94.3\%$). The Planck+WP constraints on 
$N_{\rm eff}$ and other parameters, such as the Hubble constant and the matter density, 
are in very good agreement with those obtained from the WMAP9+SPT dataset.
It is clearly worth to note that two very different datasets provide an indication
for a larger value of the effective neutrino number.
In general, we found a tension on the derived parameters between the Planck+WP and the
WMAP9+ACT datasets. This clearly indicates that the inclusion of the ACT dataset in a combined
Planck+ACT has to be carefully considered.

The anomalous lensing amplitude from Planck+WP is more consistent with the results
obtained from the WMAP9+ACT dataset, which also provide a $\sim 2 \sigma$ indication
for a larger value. However, since the same signal is not found in the trispectrum
analysis \cite{Ade:2013tyw}, the nature of this anomalous lensing amplitude needs further investigation.
Moreover, our analysis clearly demonstrates a correlation between $A_{\rm L}$ and the
main cosmological parameters.

The hints for new physics from the Planck+WP dataset are confirmed when the HST measurements are included and
are weakened when the BAO dataset is considered. The CMB+HST+BAO analysis also
suggests the presence of anomalous values but at smaller statistical significance.

It will be probably duty of the next Planck data release, with the full mission
and polarization data, to provide more precise, CMB only, constraints on the neutrino
number and the lensing amplitude and to confirm or falsify these current hints for new
physics from Planck.

\subsection*{Acknowledgements}
It is a pleasure to thank Andrea Marchini and Valentina Salvatelli for helpful discussions.

\end{document}